# Constrained ferroelectric domain orientation in $(BiFeO_3)_m(SrTiO_3)_n$ superlattice

R. Ranjith, R.V.K. Mangalam, Ph.Boullay, A. David, M.B. Lepetit, U. Lüders and W.Prellier[1]

*Laboratoire CRISMAT, CNRS UMR 6508, ENSICAEN*
*6 Bd Maréchal Juin, 14050 Caen Cedex, France*

A. Da Costa, A. Ferri and R. Desfeux

*Université d'Artois, Unité de Catalyse et de Chimie du Solide, CNRS-UMR 8181,*
*Faculté des Sciences Jean Perrin, Rue Jean Souvraz, SP 18, 62307 Lens Cedex, France*

Gy. Vincze, Zs.Radi

*Technoorg Linda Scientific Technical Development Ltd.,*
*Rozsa utca 24, 1077 Hungary*

C. Aruta

*CNR-INFM Coherentia, Dipartimento di Scienze Fisiche,*
*Complesso Universitario di Monte S. Angelo, Via Cintia - 80126 Napoli, Italy.*

Ferroelectric domains were investigated using piezoresponse force microscopy in superlattices composed of multiferroic $BiFeO_3$ and $SrTiO_3$ layers. Compared to single $BiFeO_3$ thin films, a reduction in the domains size and a suppression of the in-plane orientation of domains are observed in a superlattice of $(BiFeO_3)_4(SrTiO_3)_8$, suggesting a constrained ferroelectric domain orientation along the out-of-plane <001> direction. Such modification of domain size and orientation in $BiFeO_3$-based heterostructures could play a vital role on engineering the domains and domain wall mediated functional properties necessary for device applications.

[1] prellier@ensicaen.fr

Multiferroic materials that possess the simultaneous co-existence of ferroelectricity (FE) and magnetic ordering have been studied extensively both from fundamental and technological applications point of view.[1] However, only few materials possess both the FE and magnetic ordering at room temperature facilitating their utilization in device applications.[1] $BiFeO_3$ (BFO) is one among the extensively studied simple perovskite that exhibits multiferroic (ferroelectric and antiferromagnetic) characteristics at room temperature, with an antiferromagnetic $T_N$ around 650K and FE $T_C$ around 1105 K.[2] The possible coupling between ferroelectric and magnetic ordering is still a key issue which have triggered the study of the domains and domains walls[3] which might play an important role in future devices, given their small size as well as the fact that their location can be controlled.[4] Consequently, by analyzing the ferroelectric domains, their sizes and morphology of (001)-BFO epitaxial thin films,[3,5] the strong correlation between the antiferromagnetic and FE domains has confirmed the importance of domain walls and their influences over the magneto electric coupling.[3,5] A size and process dependent striped, mosaic and a fractal type FE domain pattern was in fact evidenced for single BFO thin films.[5,6] However, the origin of such size and process dependence of the domain pattern remains ambiguous,[5,6] despite their extensive study utilizing piezoresponse force microscope (PFM).[7,8] The FE polarization in BFO thin films is known to be along the pseudo cubic body diagonal <111>.[7,8] Hence, eight different orientations of polarization are theoretically possible along the body diagonals $\{111\}_C$ of a pseudo cubic unit cell.[7,8] The polarization variants and its switching give rise to the formation of 180°, 71° and 109° domains and their respective domain walls in BFO thin films.[7,8] The fractal domains (whose origin still remains ambiguous) are expected to arise from the crystal anisotropy and the presence of pinning defects.[5,6] Since, the formation of irregular domain walls are costly in terms of elastic energy consideration, it could arise from extrinsic factors such as process induced defects, misfit dislocations and strain fields.[5]

Recent study on the strain effect on epitaxial (001) BFO thin films shows that, though the magnitude of the polarization remains unchanged, the polarization variants in BFO could be altered by strain.[9] A strain-induced out-of-plane rotation of polarization from the (111) plane to the (100) plane was observed.[9] In addition to that strain-induced rotation of polarization[9], Lee et al.[10,11] observed the coupling of magnetic ordering with the orientation of polarization, which collectively suggests a possible way to tune the properties with an external parameter leading to a fixed orientation of polarization. In order to experience the strain-induced effect in BFO, epilayers needs to be grown with ultra small thicknesses, such that the strain would not be relaxed.[12] In addition, earlier studies on BFO show that one of the major hindrances in utilizing the BFO thin films in device applications is the leakage current at lower thicknesses.[13] Considering the reduction in leakage and maintaining a strained BFO layers, the superlattice approach was employed.[13] For these reasons, the simple perovskite $SrTiO_3$ (STO) having a good lattice matching with BFO and a larger band gap (in comparison to BFO) was selected.[13]

In this letter, superlattice structures comprising $(BiFeO_3)_m(SrTiO_3)_n$ were fabricated. FE domain patterns of these heterostructures were studied by PFM. The domain size and their corresponding in-plane (IP) and out-of-plane (OP) orientation was studied in a $(BFO)_m(STO)_n$ superlattice structure with different periodicity. Further studies on bilayers of BFO and STO were performed to confirm and understand the constraints observed in the domain orientation, and the results observed in a $(BiFeO_3)_4(SrTiO_3)_8$ superstructure are presented in this letter.

Thin films of BFO, STO and their superlattices were grown on conductive buffer layer of $LaNiO_3$ deposited on (001)-oriented STO substrates (CrysTec, Germany), at 700°C in an oxygen pressure of 20 mTorr using a multitarget pulsed laser deposition technique. The superlattice structures were synthesized by repeating the couple consisting of 'm' unit cells thick BFO layer and 'n' unit cells thick STO layer, with m,n taking integer values from 2 to 20, keeping a constant total thickness of the superlattice (1200Å). The details of the structural and macroscopic electrical measurements could be found elsewhere.[13,14] FE domain pattern of the superlattices was observed using a modified commercial atomic force microscope (Multimode, Nanoscope IIIa, Digital Instruments) operating in a piezoelectric constant mode equipped with a platinum/iridium coated silicon tip with spring constant of ~ 2.5 N/m. Frequency and amplitude of the driving AC voltage were adjusted to 6.15 kHz and 3.0 $V_{pp}$ respectively, to obtain optimized contrasted images. Out-of-plane PFM (OP-PFM) and in-plane PFM (IP-PFM) images were performed to reveal the domains orientation and the corresponding polarization.

The micro-structural quality of the superlattices was investigated using a JEOL2010F transmission electron microscope on cross-section specimens prepared on samples with or without $LaNiO_3$ buffer layer, by the standard techniques using mechanical polishing followed by ion-milling (Technoorg Linda). Fig. 1(a) shows a representative High Resolution Transmission Electron Microscope (HRTEM) image of the top part of a $(BiFeO_3)_4(SrTiO_3)_8$ superlattice structure with a periodicity of $\Lambda \approx 47$ Å ($\approx 12 \times a_{STO}$). The HRTEM image reveals the coherent growth of BFO and STO layers as confirmed by the selected area electron diffraction (SAED) pattern which displays satellite reflections close to the intense spots characteristic of a $[100]_{STO}$ zone axis pattern (insert of Fig.1a). The periodicity, calculated from the SAED pattern on various samples, matched well with the average periodicity calculated using X-ray diffraction pattern.[14] Noticeably the TEM investigations reveal the presence of strain throughout the film that could be imaged in bright field as dark contrasted zones (see top left and bottom right of Fig.1a). The presence of strains in the superlattice structure was further investigated by x-ray diffraction through the reciprocal space mapping (RSM) performed around the (103) reflection. The obtained RSM (Fig. 1b) confirms that the whole superlattice takes the same in-plane lattice parameter value of STO (3.905Å). Also, no distinctive splitting was observed in the average $SL_0$ peak as observed for BFO thin films with monoclinic symmetry due to (103) and (-103) reflections. Collective observations from the RSM and TEM lead to a calculated average out-of-plane and an in-plane sublattice parameters for the superlattice of 3.94Å and 3.905Å, respectively. This indicates an out-of-plane strain ($\varepsilon_{zz}$) of 1.39%, while the in-plane compressive strain ($\varepsilon_{xx}$ $\varepsilon_{xy}$ & $\varepsilon_{yy}$) is close 0.51%, as compared to bulk BFO (3.96Å).[2]. Such stress would be higher for the superlattice with lower period than that of higher period. Furthermore, comparative Raman spectral studies on BFO thin film with bulk ceramic sample shows a blue shift in $A_1$(TO) mode, which could be attributed to the strain in BFO thin film.[15]

The macroscopic ferroelectric polarization and the leakage behavior of the $(BFO)_m(STO)_n$ superlattice structures has been previously reported.[13,14] Further investigations of the FE domain pattern of the fabricated thin films were studied by PFM as explained earlier. Initial studies on the FE domain of a single layer BFO (120 nm) agrees well with previous results[5] for a given thickness.[16] The FE domains of $(BFO)_4(STO)_8$ superlattice are presented here, as well as BFO/STO bilayers for references. Typical morphology and grain sizes were observed for the superlattices. The $R_{rms}$ roughness was measured to be close to 1 nm. FE domain pattern was observed by mapping the piezo force experienced by the cantilever normal to the surface of the film (OP-PFM, see Fig. 2 (a)). It is conventional to assign the bright and dark regions to the domains that correspond to the upward force and downward force experienced by the cantilever respectively, and the vice versa is theoretically plausible.[3-8] The PFM images reveal the presence of both up and down oriented domains in a superlattice structure. While the domain size is different in the superlattice in comparison to the BFO single layer of same total thickness, whereas, the domain pattern of the superlattice structures are similar to those of the single layer BFO thin films.[5,16] Thus, the superlattice structures revealed clearly a reduction in the domain size in comparison to the epitaxial thin films of BFO on (001)-oriented STO.[16] The domain size was calculated from the conventional method of counting the domain walls in various directions.[17] In the case of epitaxial BFO thin films, a fractal-kind of domain pattern is widely observed in spite of the dependence of the domain pattern on the processing.[5,12,16] The domain walls of a (BFO) (STO) SL in a given scan region and given direction was counted and compared with the BFO single layer. In comparison the domain sizes of the superlattice were observed to be smaller by 30-50 nm than the BFO single layer of same total thickness. Several explanations proposed for the possible extrinsic modification of the domain size for a given total thickness and crystallite size includes the reduction of domain wall formation energy due to the strain across the superlattice structures[5] or even the strong electrostatic coupling between the layers.[18]

In the $(BFO)_4(STO)_8$ superlattice structure studied, only OP-PFM images are observed and, no domain contrasts was observed in the IP-PFM images. Fig. 2a and 2b show the OP-PFM and IP-PFM, respectively with BFO layer on top. To understand the role of the top layer (ending the superlattice structure), we also present in Fig. 2 (c and d) the OP-PFM and IP-PFM, respectively for the same $(BFO)_4(STO)_8$ superlattice ($\Lambda \sim 47$Å) with STO layer on top. The aforementioned observations are confirmed by scanning different regions and on different rotations of the sample. Consequently, the absence of contrast in IP-PFM images indicates the absence of domains, oriented in the lateral directions with respect to the cantilever tip.[19] Considering the weak thickness of the BFO layers and the crystallite size in all superlattice structures, it appears that the in-plane strain imposed by the STO substrate and intercalation layers on the BFO layers induces domains oriented perpendicular to the film surface, and thus hinders other orientations. Indeed, the very small thickness (a few unit cells only) of the BFO layers associated with first, the strong cubic character of the STO substrate, and second, the reinforcement of this character by the STO intercalated layers can be expected to impose to the BFO layers a strong tetragonal constrain. Let us remind the reader that such tetragonal geometry was recently observed in bulk ceramics[20] and in $(La_{2/3}Ca_{1/3}MnO_3)_n(BaTiO_3)_m$ superlattices.[21] In addition, recent theoretical studies provided significant evidences of domain orientation modulation by strain constraints in BFO[22] while similar modification of dipolar orientations were observed in the case of $BaTiO_3$ and STO-based superlattice structures.[23]

Let us now consider the consequences of such a geometrical constrain on the BFO layer. The STO belonging to a space group of *Pm-3m*, posses 48 symmetry operators among which, four of them involve only in-plane transformations and can thus be imposed on the BFO layers by strains at interfaces. Those symmetry operations are namely the four-fold and the two-fold rotations around the **c** axis, and the two (**a**,**c**) and (**b**,**c**) mirror planes. It is easier to see that imposing such symmetry operations (possibly associated with a doubling of the unit cell in the **a** and **b** directions) on BFO would involve smaller atomic movements. A simple symmetry calculation shows that the only polarization direction allowed by both the four-fold rotation and the two-fold rotation (which is also the combination of the two reflection operations) is the out-of plane **c** direction as observed in superlattice structures, which is is confirmed by the absence of splitting in RSM (Fig.1b).

The observed effect was further confirmed by the study of a couple of bilayer thin films $(BFO)_n(STO)_n$, with individual layer thickness kept around 65nm. Fig. 3a and 3b show the OP-PFM image and IP-PFM image of a bilayer sample with BFO on top. The domain images and the color contrasts observed in both OP-PFM and IP-PFM clearly show the presence of domains oriented both normal to the surface of the film and in the lateral directions with respect to the tip. It is worth mentioning that in a bilayer with BFO on top, the domain size, and pattern are consistent with a single layer BFO thin film of the same thickness (~ 65nm).[5,16] The good correlation of this bilayer sample (with BFO on top) with the single layer BFO thin films[6,8] suggests that the absence of in-plane orientation of domains is indeed a direct effect of the formation of a superstructure. In the case of the other bilayer sample having STO on top, a clear dark and bright contrast is observed for the OP-PFM mode (see Fig. 3c). Considering the domain size and the pattern observed, it could be a combinatorial effect of interfacial strain, induced polarization in STO and the underlying BFO layer. Figure 3(d) shows the IP-PFM of the bilayer sample with STO on top, which does not exhibit any domain contrast. The absence of in-plane orientation of domains in the bilayer with STO on top could be purely due to the constraints on polarization variants of STO-an incipient ferroelectric-considering the thickness of the STO top layer.[24] Extensive studies on the correlation of individual layer thickness, periodicity and strain in domain orientation of $(BFO)_m(STO)_n$ superlattices are currently under progress. Nevertheless, the domain images observed on the superlattice structures and the bilayers, respectively, confirm that the strain constraints introduced in an epitaxial superlattice structure of BFO and STO, grown on STO<001> substrates, drives the orientation of FE domains along the growth direction, and hinders the in-plane domain orientations. Poling experiments were also carried out at the surface of superlattices over 4 $\mu m^2$ square area by applying a – 10.5 V dc bias on the probe during scanning. As observed on the OP-PFM image (Fig. 4), black contrast evidences the existence of poled domains with downward upward polarization. No white or grey intermediate contrasts coexist in this area confirming the 180° switching of the out-of-plane upward domains.

In summary, ferroelectric domains of $(BFO)_4(STO)_8$ superlattice structures were analyzed by PFM imaging. A significant variation in domain size was observed in the superlattice structures in comparison to the BFO thin films of same total thickness. The analyses of the OP-PFM and the IP-PFM domain images of superlattices with different stacking configuration, confirmed that the interfacial strain present in the superlattice, confines the domain orientation along the direction normal to the film surface. The analysis of OP-PFM and IP-PFM of BFO and STO heterostructures suggest a possible structural modification of BFO layers in a $(BFO)_m(STO)_n$ superlattice structure. Consequently, the interfacial strain present in a superlattice structure can be utilized to externally control the density of domain walls, and the domain orientation, which could be a potential tool in terms of application of these superlattices utilizing domain orientation and domain wall, mediated functional properties.

This work was carried out in the frame the STREP MaCoMuFi (NMP3-CT-2006-033221) supported by the European Community and by the CNRS, France, the CEFIPRA and the Région Basse Normandie thought the CPER and the C-Nano program. A.F is grateful to the Nord-Pas de Calais Region for its support to carry out this work. The authors would also like to thank Prof. Ph. Ghosez, Prof. J.F. Scott and Prof. D.Chateigner for helpful discussions, Dr. L. Mechin, Mr. C. Fur, Mr. J. Lecourt, in the sample processing and L. Maës for technical support in PFM experiments.

[1] W. Prellier, M.P. Singh and P. Murugavel, J. Phys. Cond. Matter, **17,** R803 (2005).

[2] J.Wang, J.B.Neaton, H.Zheng, V.Nagarajan, S .B. Ogale, B.Liu, D.Viehland, V.Vaithyanathan, D.G.Schlom, U.V.Waghmare, N.A.Spaldin, K.M.Rabe, M.Wuttig and R.Ramesh, Science, **299,** 1719 (2003).

[3] Y-H. Chu, L.W. Martin, M.B. Holcomb, M. Gajek, S-J Han, Q He, N Balke, C-H Yang, D Lee, W. Hu, Q. Zhan, P-L Yang, A.F Rodriguez, A. Scholl, S.X. Wang and R. Ramesh, Nature Materials **7**, 478 (2008).

[4] J. Seidel, L. W. Martin, Q. He, Q. Zhan, Y.-H. Chu, A. Rother, M. E. Hawkridge, P. Maksymovych, P. Yu, M. Gajek, N. Balke, S. V. Kalinin, S. Gemming, F. Wang, G. Catalan, J. F. Scott, N. A. Spaldin, J. Orenstein & R. Ramesh, Nature Materials, **8**, 229 (2009).

[5] G. Catalan, H. Bea, S. Fusil, M. Bibes, P. Paruch, A. Barthelemy and J.F. Scott, Phys. Rev. Lett., **100,** 027602 (2008).

[6] Y.H. Chu, T. Zhao, M.P. Cruz, Q. Zhan, P.L. Yang, L.W. Martin, M. Huijben, C.H. Yang, F. Zavaliche, H. Zheng, and R. Ramesh, Appl. Phys. Lett, **90**, 252906 (2007).

[7] F. Zavaliche, R.R Das, D.M. Kim, C.B. Eom, S.Y. Yang, P. Shafer and R. Ramesh, M.P Cruz, Appl. Phys. Lett., **87,** 182912 (2005).

[8] F. Zavaliche, P. Shafer, R. Ramesh, M.P Cruz, R.R Das, D.M. Kim and C.B. Eom, Appl. Phys. Lett., **87**, 252902 (2005).

[9] H.W Jang, S.H. Baek, D. Ortiz, C.M. Folkman, R.R. Das, Y.H. Chu, P. Shafer, J.X.Zhang, S. Choudhury, V. Vaithyanathan, Y.B. Chen, D.A. Felker, M.D. Biegalski, M.S. Rzchowski, X.Q. Pan, D.G. Schlom, L.Q. Chen, R. Ramesh and C.B. Eom, Phys. Rev. Lett., **101**, 107602 (2008).

[10] S. Lee, T. Choi, W. Ratcliff II, R. Erwin, S-W. Cheong, and V. Kiryukhin, Phys. Rev. B, **78**, 100101(R) (2008).

[11] S. Lee, W. Ratcliff II, S-W. Cheong, and V. Kiryukhin, Appl. Phys. Lett., **92**, 192906 (2008).

[12] H.Bea, S. Fusil, K. Bouzehouane, M.Bibes, M. Sirena, G. Herranz, E. Jacquet, J.P Contour, A.Barthelemy, Jap. J. Appl. Phys., **45,** L187 (2006).

[13] R.Ranjith, W. Prellier, Jun Wei Cheah, J. Wang and T. Wu, Appl. Phys. Lett. **92,** 232905 (2008) and references therein.

[14] R.Ranjith, B.Kundys and W. Prellier, Appl. Phys. Lett., **91,** 222904 (2007).

[15] Y.Yang, J.Y.Sun, K.Zhu, Y.L.Liu, and L.Wan, J. Appl. Phys., **103,** 093532 (2008).

[16] R. Ranjith, U. Lüders, W. Prellier, A. Da Costa, Ida Dupont and R. Desfeux, J. Mag. Mag. Mater. **321**, 1710 (2009).

[17] Different regions in the horizontal, vertical directions and the diagonal directions of the image observed were considered and averaged to obtain the domain sizes which were calculated by counting the domain walls in different directions of the given image. Later was averaged out taking into account all the directions.

[18] A.L. Roytburd; S. Zhong and S.P. Alpay, Appl. Phys. Lett. **87**, 092902 (2005).

[19] A. Gruverman, A. Kholkin, A. Kingon and H. Tokumoto, Appl. Phys. Lett. **78**, 2751(2001).

[20] T.P.Comyn, T.Stevenson, M.Al-Jawad, S.L.Turner, R.I.Smith, W.G.Marshall, A.J.Bell and R.Cywinski, Appl.Phys.Lett., **93**, 232901 (2008).

[21] A. Sadoc, M.B. Lepetit, Ch. Simon, B. Mercey and W. Prellier, Submitted, M.P. Singh, W. Prellier, Ch. Simon and B. Raveau, Appl. Phys. Lett., **87**, 022505 (2005).

[22] L.J.Li, J.Y.Li, Y.C.Shu and J.H.Yen, Appl. Phys. Lett. **93**, 192506 (2008).

[23] A.Q. Jiang, J.F.Scott, Huibin Lu and Zhenghao Chen, J. Appl. Phys., **93**, 1180 (2003).

[24] E. Bousquet, M. Dawber, N. Stucki, C. Lichtensteiger, P. Hermet, S. Gariglio, J.M. Triscone and Ph. Ghosez, Nature **452**, 732 (2008), J.H.Haeni, P. Irvin, W.Chang, R. Uecker, P.Reiche, Y.L.Li, S.Choudhury, W.Tian, M.E. Hawley, B. Craigo, A.K.Tagantsev, X.Q. Pan, S.K.Streiffer, L.Q. Chen, S.W. Kirchoefer, J. Levy and D.G. Schlom, Nature, **430**, 758 (2004).

**Figure Captions:**

**Figure 1:** (Color online) $(BFO)_4(STO)_8$ superlattice (a) High resolution TEM image showing the regular alternation of BFO and STO layers. Inset (Right) shows the SAED pattern, corresponding to a $[100]_{STO}$ zone axis orientation. Insert (Left) shows the intensity profile from a line scan across the row of satellites attached to the $00\bar{1}_{STO}$ reflection. (b) Isointensity contour plot in logarithmic scale of the (103) reciprocal map. The logarithmic scale ranges from 2 to 4000. The Laue indices H and L are defined the lattice parameters of the $SrTiO_3$ substrate. The (103) peaks of LNO, STO substrate and average structure $SL_0$ are indicated on the map.

**Figure 2:** (Color online) (a) OP-PFM image of a $(BFO)_4(STO)_8$ superlattice structure with BFO on top, (b) corresponding IP-PFM. (c) OP-PFM image of a $(BFO)_4(STO)_8$ superlattice structure with STO on top, (d) corresponding IP-PFM. (Scanning speed for imaging: 0.4 Hz and a lock-in time constant of 3.0 ms.)

**Figure 3:** (Color online) (a) OP-PFM image of a (BFO)(STO) bilayer structure with BFO on top, (b) IP-PFM image of a (BFO)(STO) bilayer structure with BFO on top. (c) OP-PFM image of a (BFO)(STO) bilayer structure with STO on top, (d) corresponding IP-PFM of (BFO)(STO) bilayer structure with STO on top.

**Figure 4:** (Color online) Characteristic OP-PFM image of a $(BFO)_4(STO)_8$ superlattice structure when square area (2 x 2 $\mu m^2$) has been polarized (white region) by applying a –10.5 V dc bias on the probe during scanning. The scan size of the image is 5 x 5 $\mu m^2$.

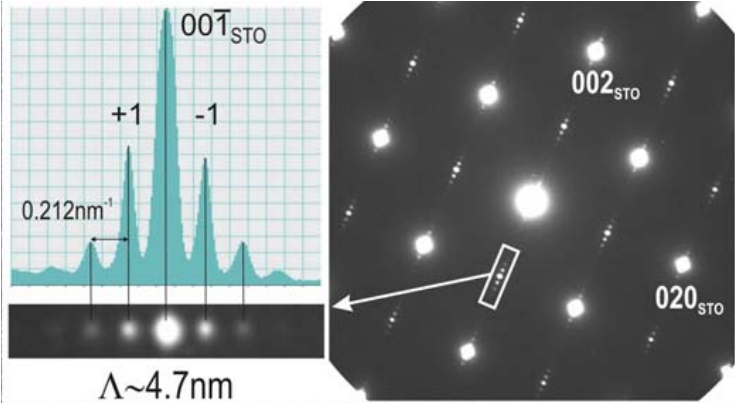
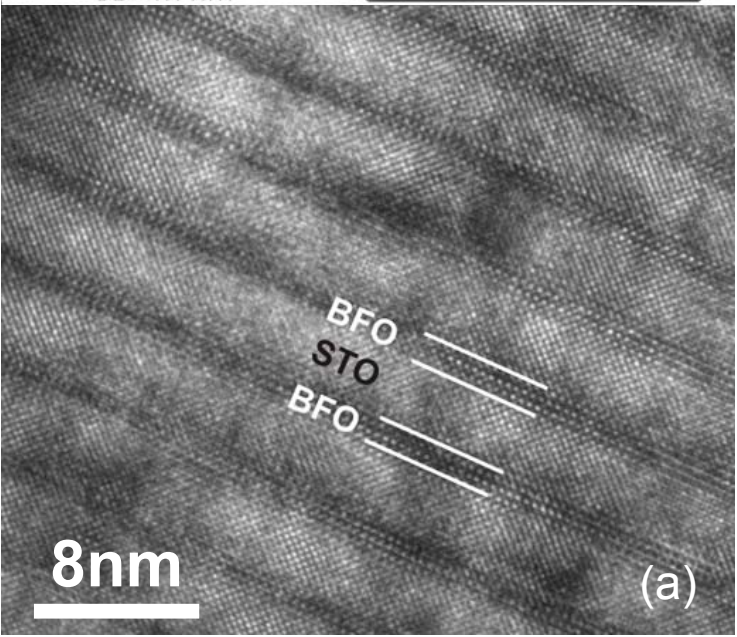
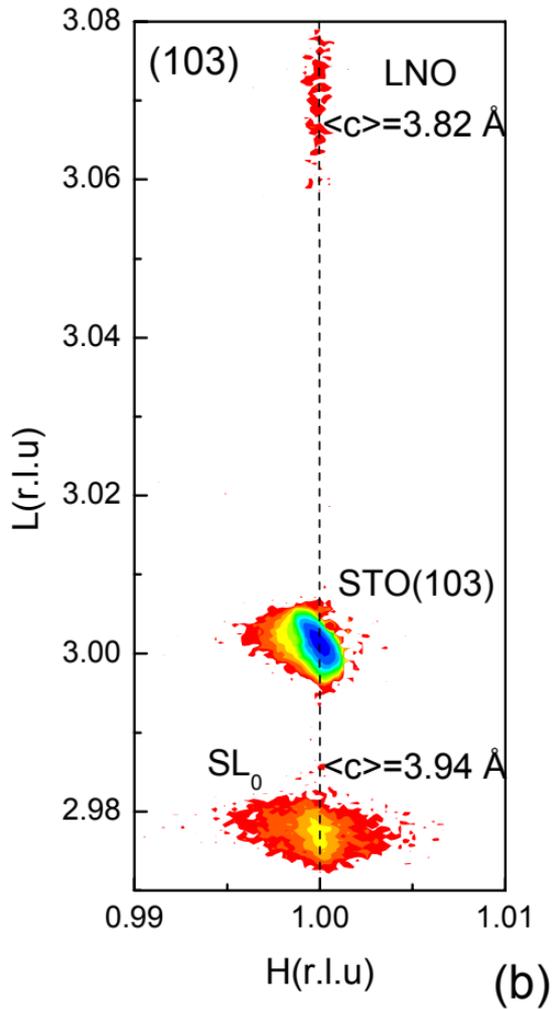

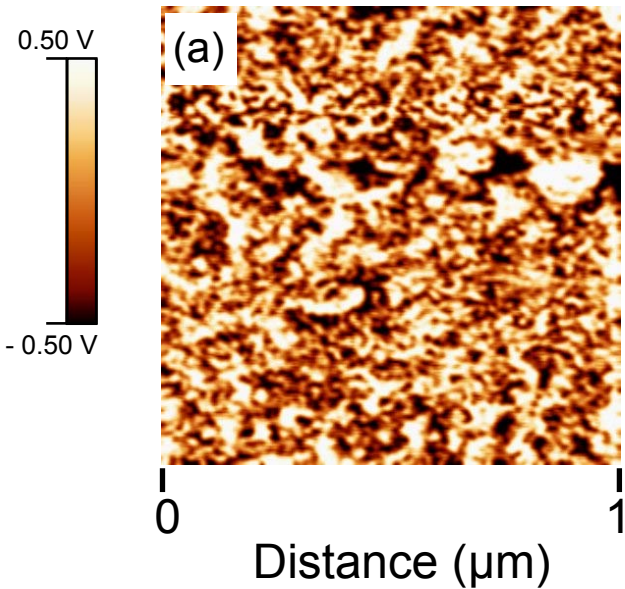 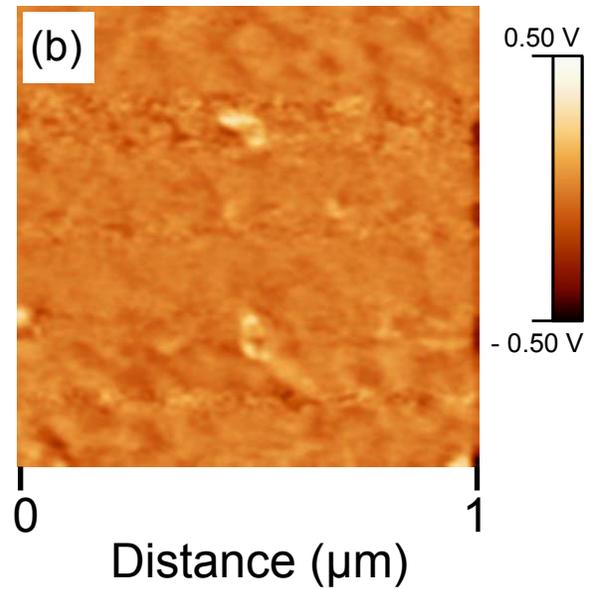
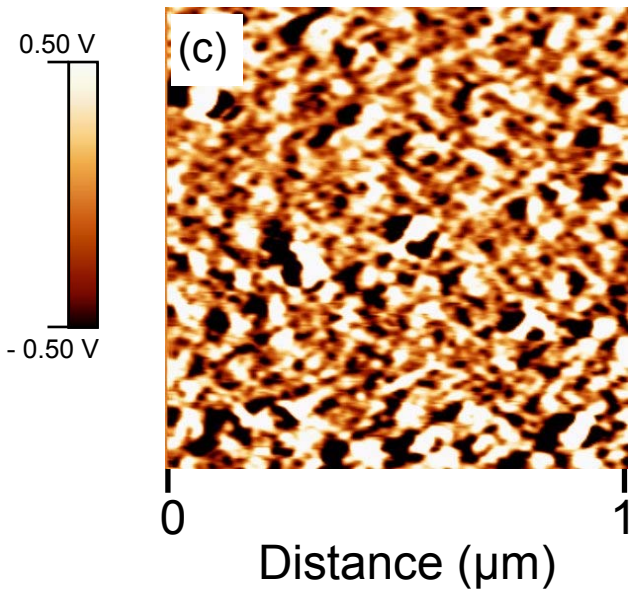 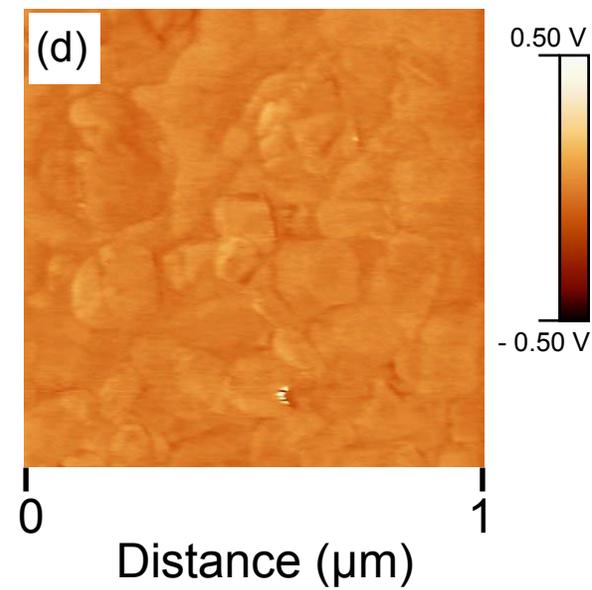

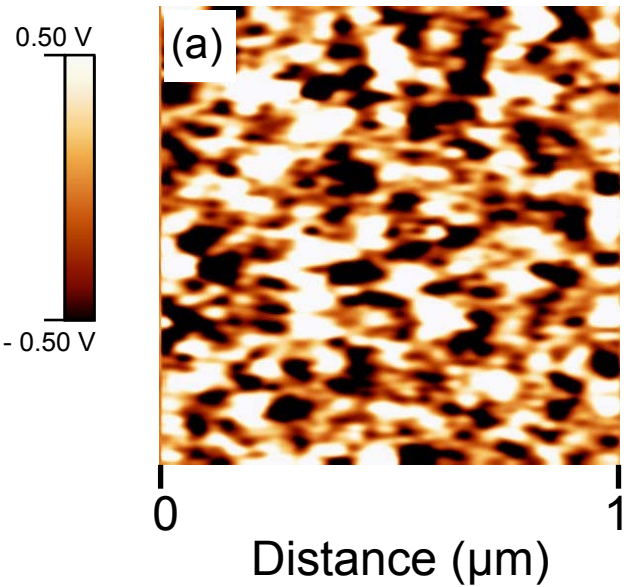 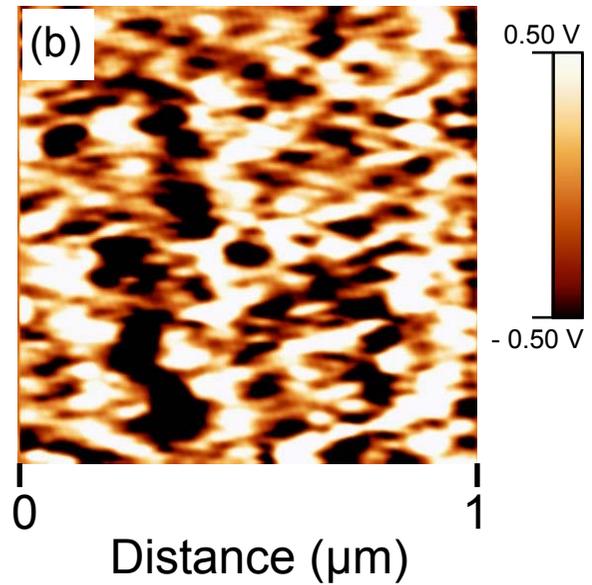
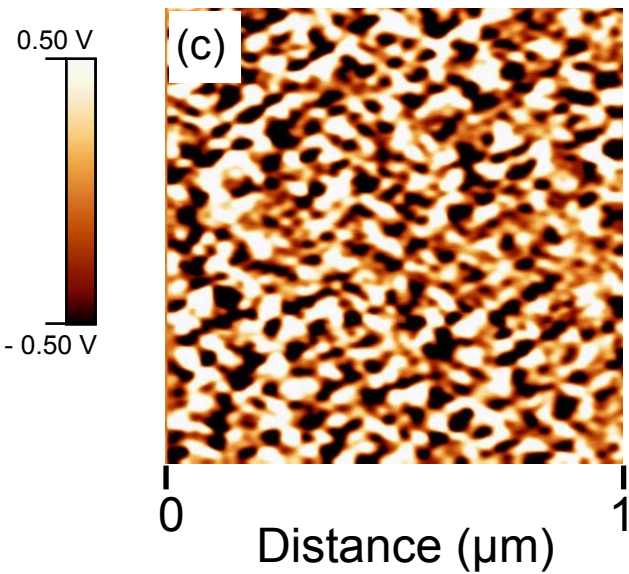 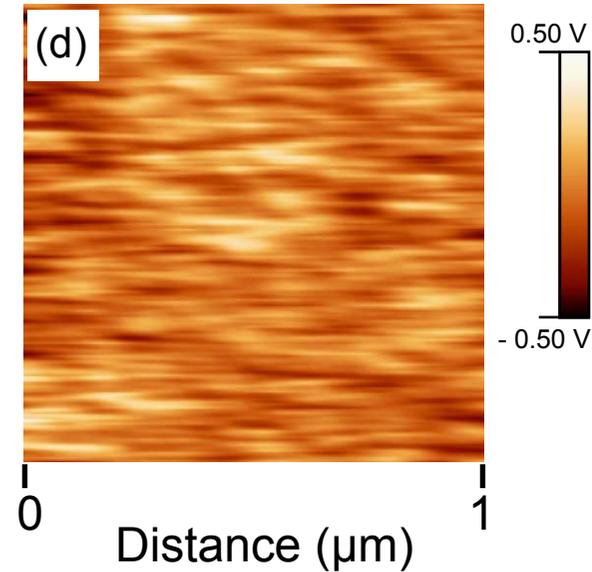

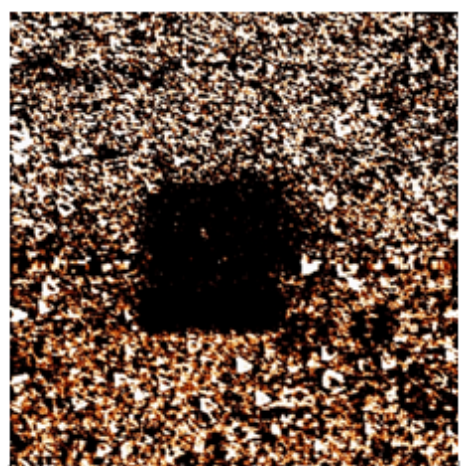